# Explore Cross-Codec Quality-Rate Convex Hulls Relation for Adaptive Streaming

Masoumeh Farhadi Nia[1, *]

1-Department of Electrical and Computer Engineering, University of Massachusetts Lowell, Lowell, MA, USA

*Corresponding author: Masoumeh_FarhadiNia@student.uml.edu

**Abstract**

With the ongoing advancement of video technology and the emergence of new video platforms, suppliers of video contents are striving to ensure that the video quality meets the desire of consumers. Accessing a limited amount of channel bandwidth, they are often looking for a novel approach to decrease the use of data and thus the required energy and cost. This study evaluates the Quality Rate performance of H.264, H.265, and VP9 codecs across resolutions (960×544, 1920×1080, 3840×2160) to optimize video quality while minimizing bitrate, crucial for energy and cost efficiency. At this approach, original videos at native resolutions were encoded, decoded, and rescaled using FFmpeg. For each resolution, encoding and decoding were performed at various quantization levels. Quality Rate (QR) curves were generated using PSNR and VMAF metric against bitrate. Convex Hull curves were then derived and mathematically modelled for each resolution. The procedure was systematically applied to H.264, H.265, and VP9 codecs. Results indicate that increasing CRF values reduce bitrate, PSNR, and VMAF, with PSNR ranging between 20-40 dB. Logarithmic polynomial modelling of convex hulls demonstrated high accuracy, with low RMSE and high R-Squared values. These findings suggest that the convex hull of one codec can predict the performance of others, aiding future content-driven prediction methodologies and enhancing adaptive streaming efficiency.

# 1. Introduction

From the invention of the first cinematograph in 1895 by Auguste and Louis Lumiere and the world's first motion picture by the Lumieres [1], to the current era of digital and mobile video, the industry has seen substantial progress. The data revolution was well on its way, but with Covid-19 [2], the world was more convinced than ever that "Digital Life is Real Life"[1]. According to estimates, the global data volume is projected to soar to 175 ZB by 2025 [3]. Video content is a significant driver of this data growth, with streaming services, social media, and online communication apps among the top consumers of high-speed internet data [4]. Given this context, optimizing video codecs to maximize quality while minimizing bitrate is crucial. This not only ensures better user experiences but also reduces energy consumption and costs. Adaptive streaming has emerged as a key technology in this regard, automatically modifying video quality according to network conditions and device capabilities. Among the widely used codecs equipped with this technology are H.264, H.265, and VP9 that are investigated in this approach. Essential concepts like data compression, adaptive and progressive streaming, video codecs, and video quality are covered in the following subsections.

## 1.1 Data Compression

As video technology advances and new platforms emerge, content providers strive to ensure video quality meets consumer expectations. Accessing a limited amount of channel bandwidth, they are often looking for a novel approach to decrease the use of data and thus the required energy and cost. Video compression resulted from data compression, source coding [5] or bit-rate reduction is one of these advantageous approaches introduced many years ago. Data compression is synonymous with the pre-transmission data encoding phase in which the data is encoded with fewer bits compared to the original video [6]; the size of the data is simply decreased to conserve space or transmission time. Two methods of compressions are lossless and lossy [7]. Lossless compression retains all original data for complete integrity, ideal for applications like financial data encoding, while lossy compression decreases file size by eliminating unnecessary data., commonly used in audio, video, and JPEG images [7].

## 1.2 Adaptive and Progressive Streaming

Due to diverse devices and varying network speeds, adaptive streaming is essential for optimal video quality. Unlike progressive streaming [8], which uses a single video file for all devices, adaptive streaming adjusts video quality based on real-time bandwidth and CPU capacity. Dynamic Optimizer (DO) [9], enhances adaptive streaming by optimizing compression across different qualities, resolutions, and shots, making it suitable for various encoders. The bitrate ladder, an adaptive-streaming concept, encodes videos into multiple streams for different resolutions and bitrates. Content-agnostic bitrate ladders, utilizing machine learning, predict optimal bitrates, reducing encoding costs and improving efficiency [10].

---

[1] Fareed Rafiq Zakaria

### 1.3 Video Codecs and Standards

A codec is an algorithm that performs its purpose whenever a program running on an operating system calls the file. It is the abbreviation of 'coder-decoder' or 'compressor-decompressor' [11]. It is also possible to embed codec into hardware, such as smartphones or cameras, to change incoming video and audio into a digital signal. At the point of capturing and playing back the product, this is applied in real time. The codec also transforms the feature into a playback format and converts digital video and audio formats. For viewing, archiving, and transferring, the program compresses the audio and video data into an achievable size. This paper examines the performance of Advanced Video Coding (H.264), High Efficiency Video Coding (HEVC or H.265), and VP9. H.264 [12], released in 2003, provides high compression for both conversational and non-conversational applications, widely used in Blu-ray and cable broadcasting. HEVC [13], finalized in 2013, offers a 50% reduction in bit-rate compared to H.264, utilizing hybrid coding methods for superior video quality. VP9 [14], introduced by Google in 2013 as part of the WebM project, also achieves a 50% bitrate reduction while maintaining quality through efficient compression and motion vector prediction.

### 1.4 Video Quality

Quality of Experience (QoE) [15] broadly illustrates the user-oriented quality of video, influenced by factors such as viewing setup, display properties, individual viewer preferences, and interaction with the service. Video quality can be evaluated through subjective and objective measures. Subjective assessments rely on human evaluation, while objective assessments use mathematical algorithms. Picture metrics and reference information are critical categories for objective assessments. In our research, we utilize Peak Signal to Noise Ratio (PSNR) for its simplicity and which is the prominent metric in video codec comparison studies.

## 2. Related Work

While previous studies have compared the performance of mentioned codecs from various perspectives, this research takes a novel approach by examining the similarities between them. Specifically, it models the Quality Rate (QR) convex hull curves across different resolutions (960×544, 1920×1080, 3840×2160) to explore how well one codec's convex hull can predict another's performance.

Previous studies have focused on dynamically optimized adaptive streaming, such as MPEG DASH [16], to satisfy the growing demand for video content online. These methods adapt video streams based on receiver characteristics like screen size and internet speed, enhancing video quality and user satisfaction. Projects [17], [18], and [19] have been designed to optimize the QR quality features dynamically by generating QR convex hull curves assessed in VMAF (Video Multi-Method Assessment Fusion) and PSNR (Peak Signal-to-Noise Ratio). Recent studies such as Yin et al. [20]and Lekharu et al. [21] have further refined these methodologies by integrating machine learning models to predict streaming performance, demonstrating improved accuracy and efficiency.

Several experiments have been conducted before our approach. For instance, predicting the bitrate for adaptive streaming [22], subjective comparison of HEVC and AV1 for adaptive streaming [23], and analysis of H.264, H.265/ HEVC, and VP9 coding efficiency with a focus on BD (Bjøntegaard Delta) curves comparison scaled in PSNR. Some studies, such as [24], emphasized the time required for performance completion, suggesting codec preferences for real-time and non-real-time video streaming. Additionally, [25], examined the compression performance of various codecs, highlighting BD-Rate savings for on-demand videos to determine the most efficient codec. Recent works by Valiandi et al. [26] Saha et al. [27], and Yang et al [28], valuating five codecs for volumetric video compression efficiency and quality, have expanded on these analyses, incorporating more extensive datasets and advanced statistical methods to enhance the reliability of their findings.

However, most previous works analysed codecs comparatively without considering the potential correlation between the performance of different codecs. Our study addresses this gap by investigating the relationship between the convex hulls of different codecs for the same content. This approach aims to develop a prediction methodology, allowing the use of one codec to predict the quality performance of others. By encoding a specific codec, like HEVC, we can use the identified correlation to predict the expected QR performance and convex hull of other codecs for the same content.

Our research focuses on obtaining and studying the convex hulls of QR curves at different spatial resolutions for H.264, H.265, and VP9. We aim to model the underlying relations of rate-quality performance and convex hulls between codecs for various contents. This correlation is expected to provide a content-agnostic methodology for adaptive streaming, significantly reducing the computational effort required to build bitrate ladders for different coding technologies.

## 3. Methodology

To demonstrate codec efficiency in compression, the key challenge is identifying optimal compression parameters for a desired outcome. The proposed solution constructs a mathematical model to predict compression results without actual compression, allowing quick estimation with different parameters. "Quality" is the performance metric, with PSNR and VMAF deemed most suitable. The methodology involves encoding and decoding videos at various resolutions and quantization levels utilizing FFmpeg.

PSNR and VMAF are then calculated between original and processed videos to generate quality curves for resolutions 960x540, 1920x1080, and 3840x2160. Convex hulls are extracted and modelled to calculate correlations, with all videos initially downscaled and upscaled for consistent comparison.

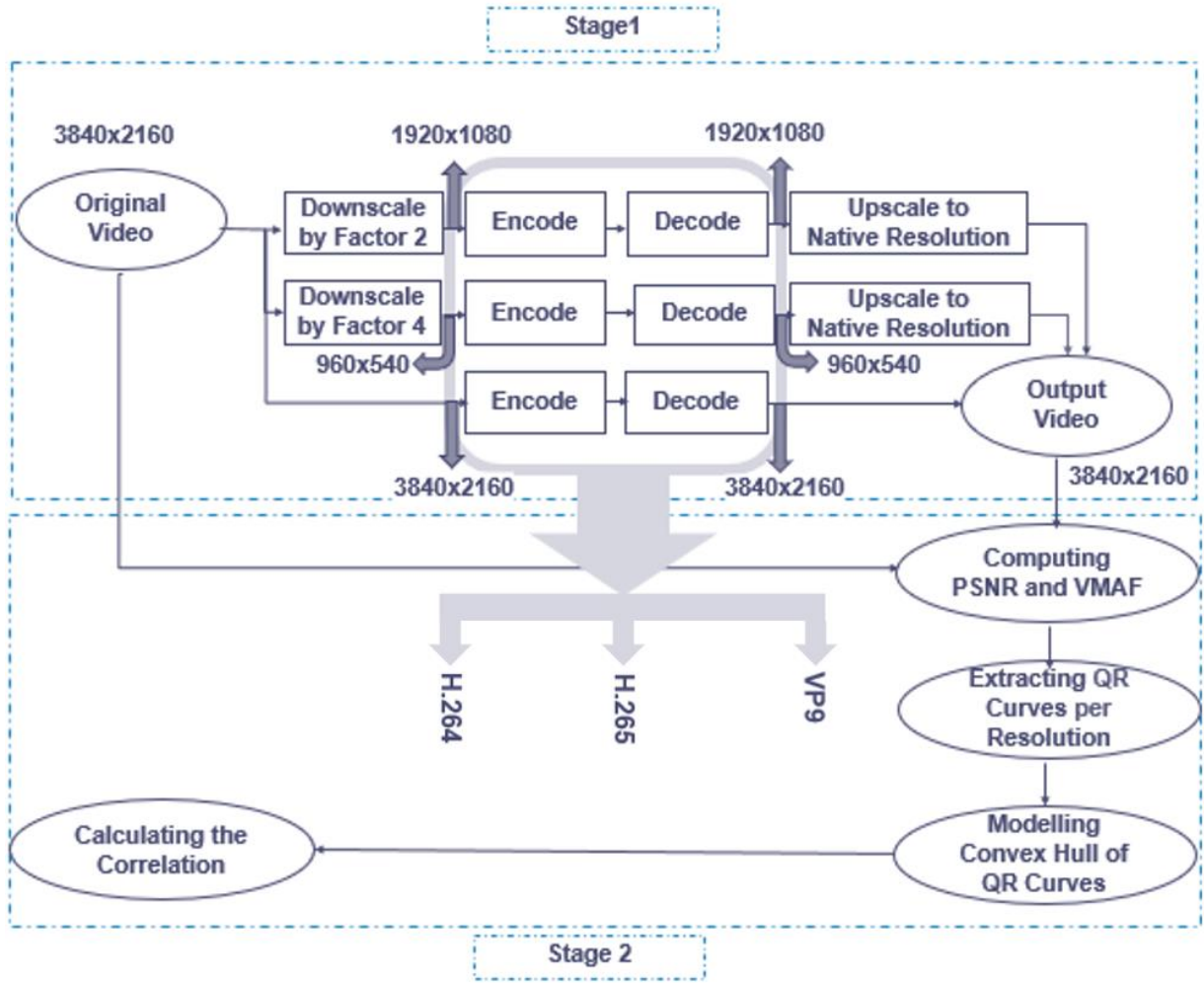

Figure 1: Flow Chart of the Research Methodology

## 3.1 Dataset Creation

The core of our research involves data creation, producing output videos to establish state-of-the-art (SOTA) quality-rate convex hulls and modeling them mathematically. This process which is memory and time-intensive, includes:

- Selecting sequences with diverse spatiotemporal characteristics (e.g., color, motion, content) representative of broadcast content.
- Utilizing a multimedia framework for resizing (downscale/upscale) with specific filters and encoding/decoding the original videos.

- Identifying the best filters for resizing.
- Determining the optimal encoding and decoding techniques for the target codecs (H.264, H.265, and VP9).

Results of our analyses for these steps are detailed in subsequent sections.

**3.1.1 Video Sequence Selections**

The dataset [29] used in this research is published by the University of Bristol and consists of nine diverse source sequences selected for their representativeness of broadcast content. These sequences vary in spatiotemporal parameters such as color, motion, and content, which are essential for measuring compression efficiency and quality metrics. The dataset is suitable for comparing quality-rate (QR) curves of different codecs (H.264, H.265, and VP9) across varied video contents, ensuring applicability to real-life viewing experiences. Source sequences of the dataset is brought in Figure 2.

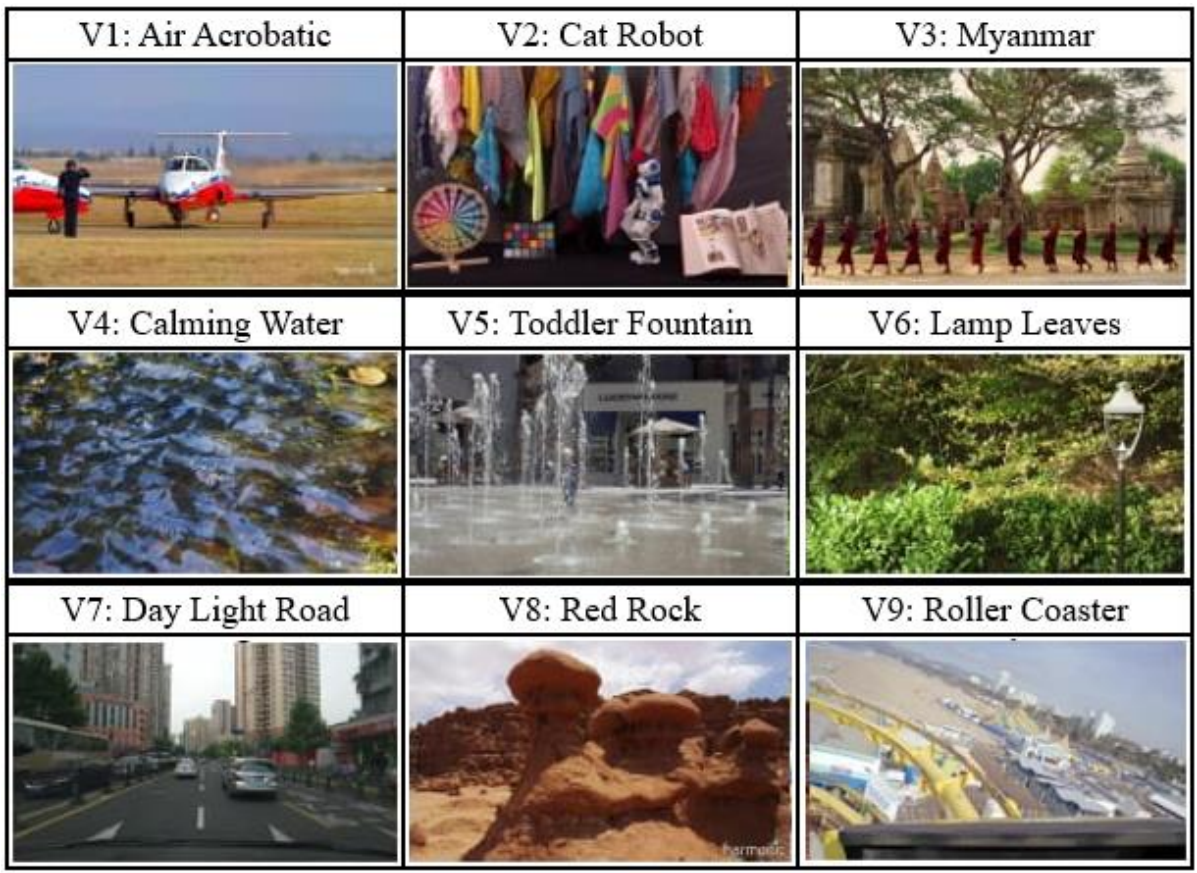

Figure 2: Source Sequences of Dataset

### 3.1.2 Content Description by Benefitting from Spatial and Temporal Information

Modern AI projects often involve semantic understanding and video content modification, which begins with video content detection and understanding. Our approach uses spatiotemporal features for content description. Temporal features (TI) measure frame differences in time domain using Equations (1-3), while spatial features (SI) assess edge movement and video complexity using Equations (4-5). These metrics categorize videos based on related behaviours.

$$M_n(i,j) = F_n(i,j) - F_{n-1}(i,j) \quad (1)$$
$$TI_n = std(M_n) \quad (2)$$
$$TI = mean(TI_n) \quad (3)$$

$F_n$ is the nth frame in the video and $(i,j)$ is the position of the pixels. $M_n$ is the difference between two adjacent frames, and the temporal information of the nth frame is determined for all pixels using the standard deviation of $M_n$. TI is determined by obtaining the mean for every frame of the video sequence.

$$SI_n = std(Sobel(F_n)) \quad (4)$$
$$SI = mean(SI_n) \quad (5)$$

$SI_n$ reflects the SI value for a given frame. It is determined by applying the Sobel filter to each pixel in the frame. Then, the standard deviation is taken in the frame. Total *SI* is computed by applying the mean function to every *SI* values of the frame [30].

### 3.1.3 Detecting a Multimedia Platform to Implement the Methodology

To simulate video telecommunication techniques, we used Ffmpeg [31], a versatile multimedia framework for downscaling, upscaling, encoding, and decoding. FFmpeg supports various formats and is highly portable across different systems. We installed the latest stable FFmpeg version on Windows 10 in June 2020. Quality metrics extraction was also implemented using FFmpeg.

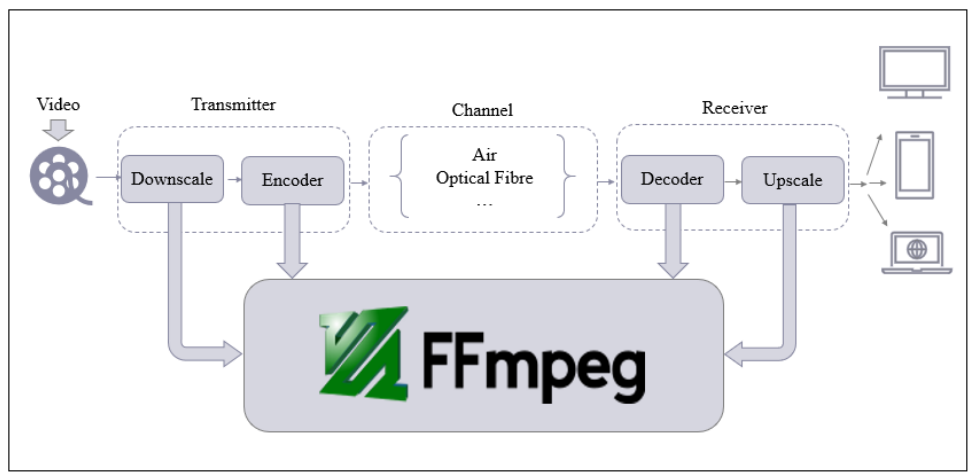

Figure 3: Utilized Platform for Scale- changing, Encoding, and Decoding.

### 3.1.4 Filter Selection for Scale-changing

Our experiment encodes and decodes videos at 960x540, 1920x1080, and 3840x2160 resolutions. Original 3840x2160 videos are downscaled and upscaled using Bicubic and Lanczos-3 filters. After evaluation, Lanczos was chosen for its superior edge preservation and consistent quality, as Bicubic resulted in blurry edges as it is shown in Figure 4.

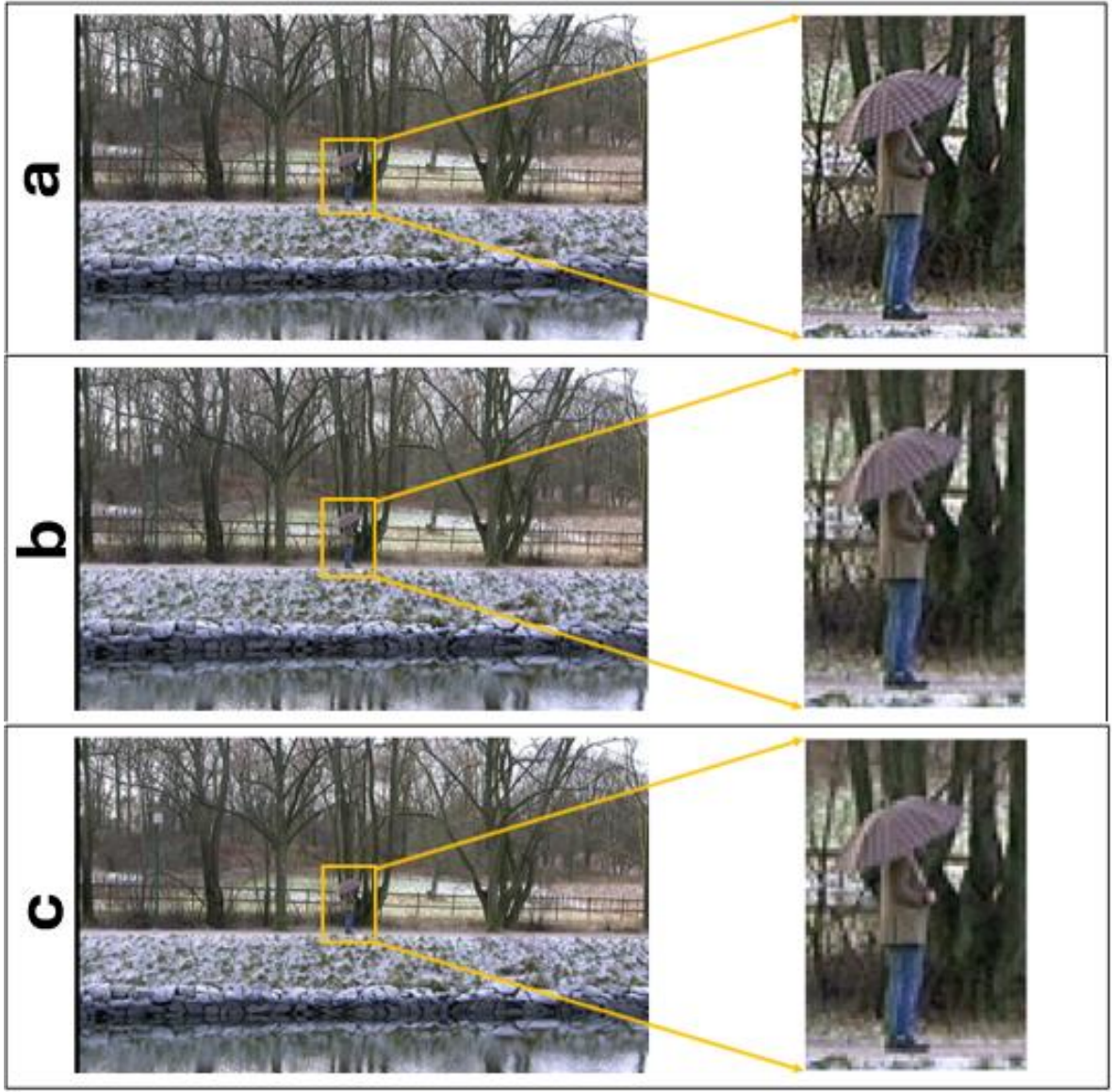

Figure 4: a) A Frame of the Original Video, b) A Frame of the Downscaled Video Using Lanczos Filter, c) A Frame of the Downscaled Video Using Bicubic Filter (All Captured at 00:00:09)

### 3.2 Compression Process for the SOTA Codecs (H.264, H.265, and VP9)

In this project, we aim to obtain PSNR and VMAF curves for specific video sequences. These curves are plotted on a two-dimensional graph with bitrate on the x-axis and quality (PSNR or VMAF) on the y-axis.

Quality is assessed by comparing the original video to the compressed output in YUV format. The steps for processing and compressing the videos using FFmpeg are as follows:

First, we convert the original format of the video sequences in the dataset to YUV. MP4, known for its suitability for internet streaming, efficient compression, and compatibility with popular media players, is not ideal for this initial step. Therefore, the original MP4 format is converted to YUV, which is like RGB but includes a luminance component for better human perception and improved error handling and bandwidth efficiency. This conversion is executed in FFmpeg using a command line.

With the videos in YUV format, we proceed to downscale them to the desired resolutions if required. The selected resolutions are 960x540, 1920x1080, and 3840x2160. The downscaling is done using specific filters to ensure quality retention. Next, we encode the video at various quantization levels, ranging from 5 to 50 in steps of 5. This encoding process is followed by decoding the encoded video. If the video was downscaled earlier, it is then upscaled back to the original resolution. This entire process is designed to be implemented across the three resolutions, with additional horizontal blocks for more varied resolutions.

Two variables, codecs and quantization parameters, change during the video process. The codecs used in this research are H.264, H.265, and VP9.

### 3.2.1 Selection of the Compression Parameter

Quantization levels, key to our approach, determine the best compression parameters through the evaluation of Quantization Parameter (QP) and Constant Rate Factor (CRF). Block-based hybrid video encoding, which is inherently lossy, uses QP to balance spatial information preservation and bitrate, with lower QP values retaining more information and higher QP values increasing distortion. CRF, the default quality setting for H.264, H.265, and VP9, ranges from 0 to 51 for H.264/H.265 and 0 to 63 for VP9. Lower CRF values yield higher quality at larger sizes, while higher values enhance compression but can reduce quality. CRF is preferred for its efficiency and quality maintenance, adjusting QP based on motion to align with human visual perception, making it suitable for offline file access and encoding MP4 format videos.

### 3.2.2 Compression Process Across Different Resolutions

For compression at 3840x2160 (4K UHD), the encoding process involves input videos in YUV format and output in MP4 format, with a CRF range from 5 to 50 in steps of 5. Specialized syntax is used for each codec from the FFmpeg Vcodec library. The decoding process involves input videos in MP4 format and output in YUV format. For compression at 1920x1080 (Full HD, FHD), the process includes downscaling the input video from 3840x2160 to 1920x1080 using the Lanczos filter. Encoding and decoding are similar to the UHD process, followed by upscaling the video back to 3840x2160 using the Lanczos filter. For compression at 960x540 (Quarter HD, QHD), the input video is downscaled from 3840x2160 to 960x544 using the Lanczos filter. Encoding and decoding follow the same process as UHD, and the video is upscaled back to 3840x2160 using the Lanczos filter. This process is demonstrated in Figure 5.

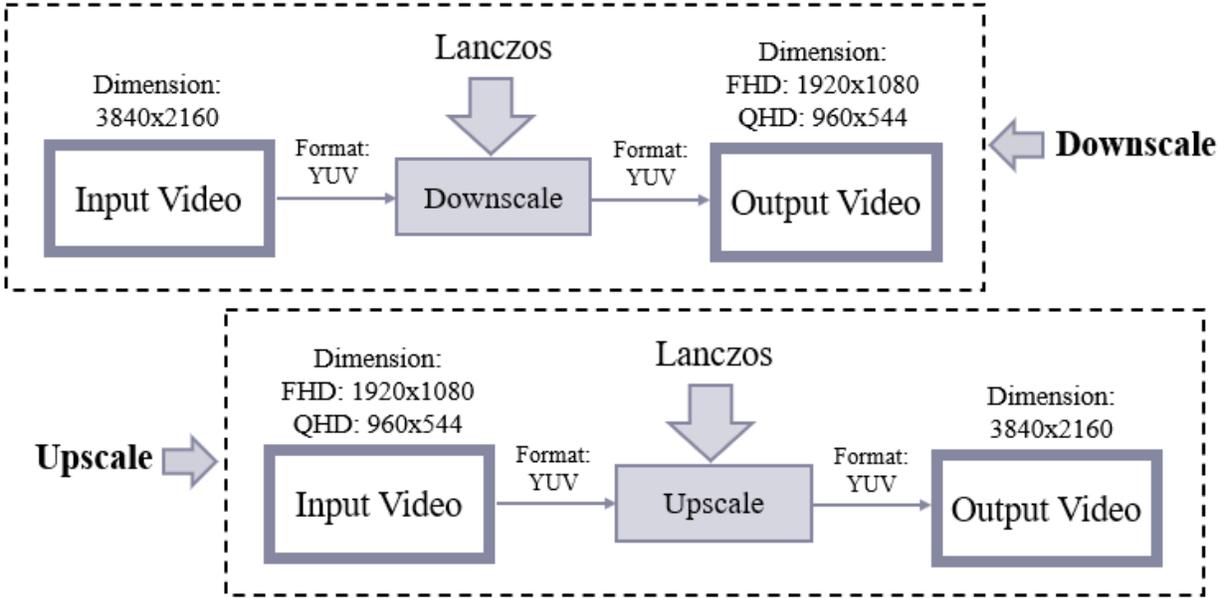

Figure 5: Dimension and Format of the Input and Output Video of the Downscale and Upscale Phase

## 3.3 Extracting QR Curves

PSNR and VMAF are two quality metrics which are extracted in our approach. In the following sections, firstly, PSNR and secondly, VMAF is investigated from the statistical viewpoint. However, before going on to the details of PSNR and VMAF calculation, we need to explain the method in which video bitrate is determined.

### 3.3.1 Obtaining Bitrate

Our goal is to plot bitrate-quality curves using PSNR/VMAF. Quality is obtained at different CRFs, and we calculate equivalent bitrates for specific PSNR values. Bitrate represents the number of video bits transmitted per second, measured in bps, Kbps, or Mbps. Bitrate calculation is outlined in Equation 6.

$$\text{Bitrate} = \frac{\text{Video Size}}{\text{Duration (number of minutes} * .0075)} \quad (6) \ [32]$$

Two methods for extracting video bitrate are, using FFprobe to analyse the encoded video and generate a text file with the bitrate value or calculating the bitrate from the encoded video's file size, frame count, and framerate, which can be implemented using MATLAB. FFprobe is utilized in this research.

### 3.3.2 Obtaining PSNR

Video coding companies extensively use PSNR as an objective video quality assessment metric, calculated by measuring the Mean Square Error (MSE) between original and decoded video sequences, [33]

$$\text{MSE} = \left(\frac{\sum_{i=0}^{M-1}\sum_{j=0}^{N-1}(S(I,j)-S_d(I,j))^2}{M*N}\right) \quad (7)$$

In the above equation, $S(I, j)$ and $S_d(I, j)$ are indicative of the original and decoded signal, respectively. The frame dimensions, meaning width and height, are defined by M and N, respectively. PSNR is then derived using Equation 8.

$$\text{PSNR}_k = 10 \log \frac{(2^b-1)^2}{\text{MSE}} \quad (8)$$

For YUV format, overall PSNR is:

$$\text{PSNR}_{4:2:0} = 16\,(4\,\text{PSNR}_Y + \text{PSNR}_U + \text{PSNR}_V) \quad (9)$$

Initially using MATLAB for PSNR calculation, we later adopted FFmpeg for faster processing.

### 3.3.3 Obtaining VMAF

For this research, VMAF [34] is used as an objective video quality assessment metric that compare the similarity between decoded videos and the original, with implementation based on methodologies from Netflix and additional adaptations for Windows.

### 3.3.4 Model QR Curves Mathematically

By using MATLAB's Curve Fitting app, we model the bitrate-PSNR / VMAF data points mathematically, generating equations to simplify codec performance comparison.

## 3.4 Extracting Convex Hull Curves and Their Mathematical Models for PSNR

Finally, we use MATLAB to obtain and model the convex hull for each curve, allowing us to compare video quality across codecs and find correlations. This involves fitting the convex hulls using polynomial functions, assessing the model's accuracy with RMSE and $R^2$ metrics.

## 4. Results

To provide some statistical vision, a limited portion of the data generated in section 3.3, UHD resolution of Myanmar video sequence is presented in Table 1 for each codec. This table is followed by an evaluation of bitrate, PSNR, and VMAF variables according to the CRF changes. In fact, we are analysing the values of these variables for any specific CRF and looking to discover the general trend of the changes in their values to see by increasing the CRF values, their trend is decremental or incremental.

Table 1: CRF, Bitrate, PSNR, and VMAF Values for H.264, H.265, and VP9 respectively.

| H.264 | H.265 | VP9 |
| --- | --- | --- |

| CRF | Bitrate (kb/s) | PSNR | VMAF | Bitrate (kb/s) | PSNR | VMAF | Bitrate (kb/s) | PSNR | VMAF |
| --- | --- | --- | --- | --- | --- | --- | --- | --- | --- |
| 5 | 695488 | 52.9669 | 92.3797 | 520521 | 49.3882 | 92.5157 | 796086 | 56.6072 | 92.4052 |
| 10 | 485177 | 48.6550 | 92.2235 | 290477 | 45.4326 | 92.3293 | 474695 | 49.8850 | 92.2808 |
| 15 | 244833 | 43.3468 | 91.6983 | 159696 | 42.4395 | 91.7173 | 286001 | 44.6691 | 91.7851 |
| 20 | 100994 | 38.6366 | 90.3284 | 81215 | 39.2211 | 90.4906 | 198634 | 42.3788 | 91.3672 |
| 25 | 43389 | 34.6996 | 87.4167 | 39267 | 35.7703 | 88.0302 | 143189 | 40.8018 | 90.8633 |
| 30 | 19060 | 31.2517 | 81.7517 | 18171 | 32.4033 | 83.1556 | 98636 | 39.0406 | 90.1058 |
| 35 | 8608 | 28.1138 | 72.5276 | 8092 | 29.2229 | 74.6307 | 63284 | 37.0052 | 88.841 |
| 40 | 4077 | 25.2584 | 60.3204 | 3198 | 26.1305 | 61.7095 | 41709 | 35.2291 | 87.2773 |
| 45 | 2008 | 22.6260 | 43.6877 | 1049 | 23.1816 | 46.0851 | 26675 | 33.4117 | 84.9734 |
| 50 | 1587 | 21.5243 | 36.3294 | 498 | 21.5127 | 38.8556 | 17940 | 31.8526 | 65.8117 |

As it is demonstrated by actual evidence and we expected owing to our literature review, by increasing CRFs, bitrate, PSNR, and VMAF are decreasing. For any step of 5-unit increase in the CRF, bitrate, experiences about 50 percent reduction in its value. PSNR and VMAF have also a decremental trend when the CRF values are Increasing.

## 4.1 PSNR and VMAF Curves of This Approach

In this section, the results of the methodology explained in chapter 3 is prepared. Firstly, the Bitrate-PSNR curves and secondly, the Bitrate-VMAF curves are provided. Figure 4. provides the legend utilized for the whole curves clearly.

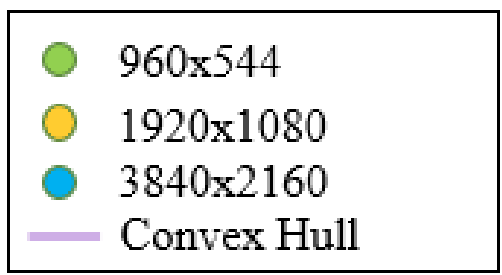

Figure 6: The Utilized Legend in the Curves

## 4.2 PSNR and VMAF Curves and their evaluation

In this section the results of the QR curves are presented.

### 4.2.1 Bitrate PSNR Curves

The Bitrate PSNR curves for nine video sequences across three different resolutions and their convex hulls are summarized in Figure 7.

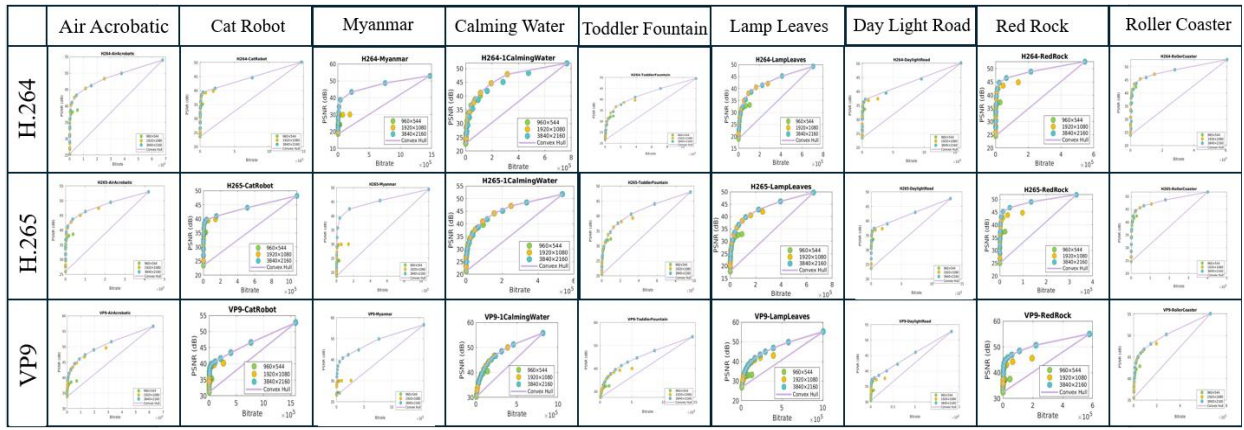

Figure 7: PSNR- Bitrate Curves for the 9 Video Sequences in the Dataset Utilizing 3 codecs (H.264, H.265, VP9)

### 4.2.2 Bitrate VMAF Curves

The Bitrate VMAF curves for nine video sequences across three different resolutions and their convex hulls are summarized in Figure 8.

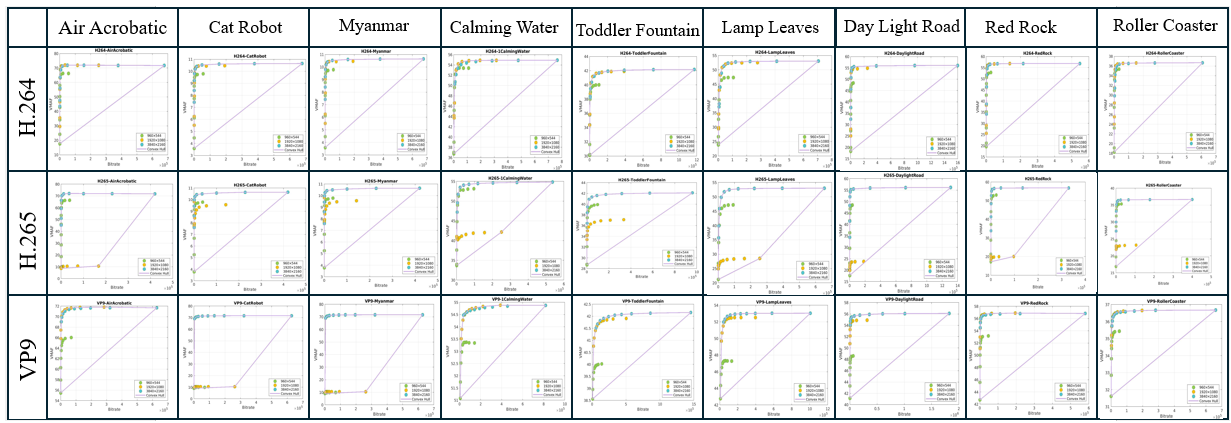

Figure 8: VMAF- Bitrate Curves for the 9 Video Sequences in the Dataset Utilizing 3 codecs (H.264, H.265, VP9)

### 4.3 Mathematical models of the PSNR Convex Hulls

By having the convex hull of the Bitrate-PSNR curves, the last step of the project which is modelling the convex hulls mathematically is implemented and the results for three codecs are provided below. For each codec, the points constructing the convex hulls of nine videos are put together and modelled. The number of points is presented in Table 2. Figures 4.3, 4.4, and 4.5 are demonstrating the modelled curves for H.264, H.265, and VP9, respectively. For each modelled curve, the related polynomial modelled curve with its accuracy is provided as well.

Table 2:Convex Hulls Points for Each Codec.

| H.264 | H.265 | VP9 |
|---|---|---|
| 129 | 139 | 135 |

**4.3.1 H.264**

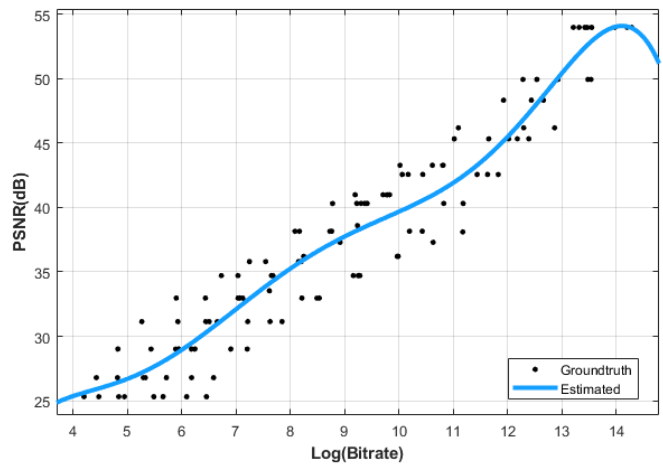

Figure 9: The Modelled Curves for Convex Hull of H.264

Linear model Poly6: (It is worth mentioning that we have considered the bitrate logarithmic which means that in the following equation x is replaced with (log x).)

$f(x) = p1*x^6 + p2*x^5 + p3*x^4 + p4*x^3 + p5*x^2 + p6*x + p7$

Coefficients (with 95% confidence bounds):

p1 = -0.0007881 (-0.002044, 0.0004681)

p2 = 0.04001 (-0.02975, 0.1098)

p3 = -0.8077 (-2.382, 0.7671)

p4 = 8.251 (-10.21, 26.71)

p5 = -44.67 (-163, 73.7)

p6 =123.2 (-269.8, 516.1)

p7 = -111.5 (-638.7, 415.6)

Therefore, the convex hull of H.264 is modelled as $6^{th}$ order polynomial.

### 4.3.2 H.265

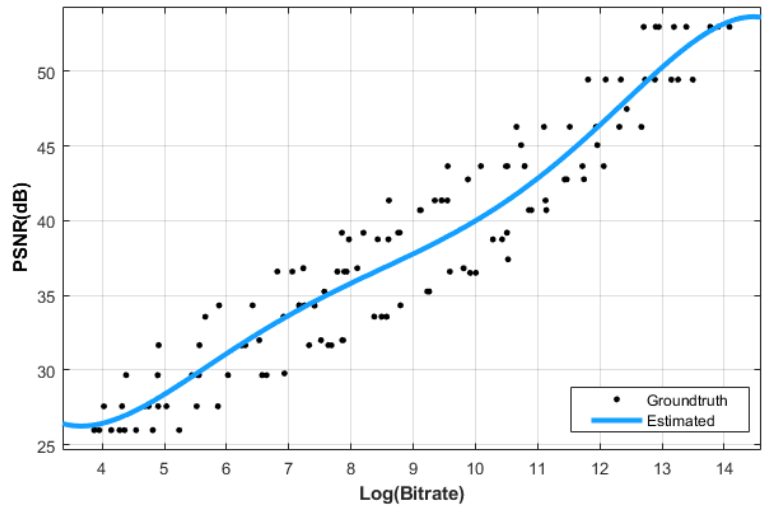

Figure 10: The Modelled Curves for Convex Hull of H.265

Linear model Poly5: (It is worth mentioning that we have considered the bitrate logarithmic which means that in the following equation x is replaced with (log x).)

f(x) = p1*x^5 + p2*x^4 + p3*x^3 + p4*x^2 + p5*x + p6

Coefficients (with 95% confidence bounds):

p1 = -0.002066 (-0.005009, 0.0008759)

p2 = 0.09133 (-0.0403, 0.2229)

p3 = -1.536 (-3.815, 0.7438)

p4 = 12.27 (-6.772, 31.31)

p5 = -44.14 (-120.6, 32.32)

p6 = 83.72 (-34.02, 201.4)

Therefore, the convex hull of H.265 is modelled as $5^{th}$ order polynomial.

### 4.3.3 VP9

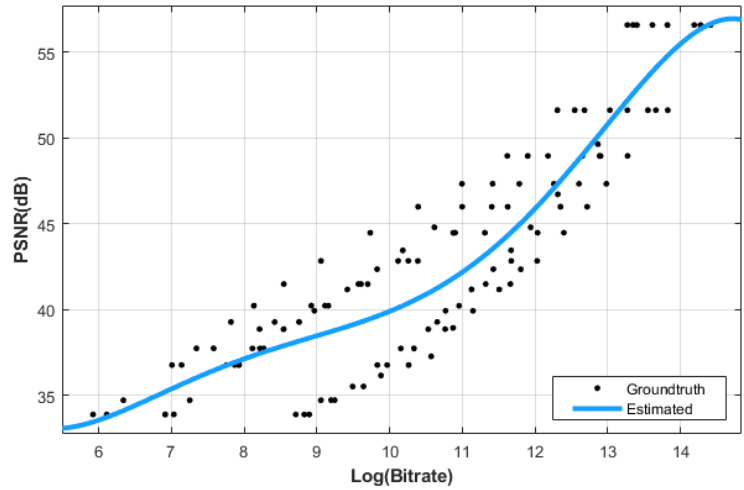

Figure 11: The Modelled Curves for Convex Hull of VP9

Linear model Poly5: (It is worth mentioning that we have considered the bitrate logarithmic which means that in the following equation x is replaced with (log x).)

$f(x) = p1*\log(x)^5 + p2*\log(x)^4 + p3*\log(x)^3 + p4*\log(x)^2 + p5*\log(x) + p6$

where x is normalized by mean 10.35 and std 2.132

Coefficients (with 95% confidence bounds):

p1 = -0.1925 (-0.5627, 0.1778)

p2 = -0.3535 (-0.7647, 0.05772)

p3 = 1.113 (-0.5089, 2.734)

p4 = 2.601 (1.254, 3.947)

p5 = 4.443 (2.879, 6.007)

p6 = 40.56 (39.81, 41.3)

As it is demonstrated above H.264 is modelled as a $6^{th}$ order polynomial and H.265 and VP9 are modelled as $5^{th}$ order polynomial. The reason is that we were changing the degree from 1 to 8 to find the best accuracy. Table 2 is exhibiting the RMSE and R-square values and our final choice for each codec that has a lower RMSE and higher R-Squared.

Table 3: RMSE and R-Squared Values for Each Degree in Polynomial Modelling

| Encoding | H264 | | H265 | | VP9 | |
|---|---|---|---|---|---|---|
| Poly order | RMSE | R-Squared | RMSE | R-Squared | RMSE | R-Squared |
| 1 | 2.421 | 0.9095 | 2.455 | 0.9 | 3.054 | 0.7732 |
| 2 | 2.368 | 0.9141 | 2.387 | 0.9062 | 2.64 | 0.8319 |
| 3 | 2.346 | 0.9164 | 2.365 | 0.9086 | 2.616 | 0.8361 |
| 4 | 2.348 | 0.9169 | 2.373 | 0.9086 | 2.606 | 0.8386 |
| 5 | 2.31 | 0.9202 | **2.365** | **0.9099** | **2.605** | **0.8399** |
| 6 | **2.305** | **0.9212** | 2.366 | 0.9065 | 0.61 | 0.8406 |
| 7 | 2.313 | 0.9167 | 2.374 | 0.9106 | 2.611 | 0.8417 |
| 8 | 2.322 | 0.9213 | 2.371 | 0.9061 | 2.621 | 0.8418 |

As it is demonstrated, the highlighted values, have a lower RMSE and higher R-Squared which resulted to a more accurate modelled curve.

## 5. Conclusion and Future Work

The video communication community is continually enhancing data transmission efficiency at lower data rates. Adaptive streaming addresses streaming concerns like speed and buffering, using codecs such as H.264, H.265, and VP9 which vary in performance and quality. This project explored and modelled the convex hull curves of these codecs. A diverse dataset was encoded and decoded in the FFmpeg environment, with QR metrics like PSNR and VMAF obtained. The results show that increasing CRF values decreases bitrate, PSNR, and VMAF. H.264's convex hull is modelled as a 6th order logarithmic polynomial, while H.265 and VP9 are modelled as 5th order logarithmic polynomials. This research aids in finding correlations between these codecs' convex hull curves, identifying codecs with high quality and lower bitrate as optimal choices.

### 5.1 Future Work

Future steps include increasing quantization levels, expanding resolution ranges, and using more videos for machine learning driven approach. In that case, we will have a predictive model for the convex hulls of the Quality Rate curves for each codec. Also, the method can be applied to more codecs.